\documentclass[conference]{IEEEtran}

\usepackage{booktabs} 
\usepackage[flushleft]{threeparttable}
\usepackage{xcolor}
\usepackage[utf8]{inputenc}
\usepackage{float}
\usepackage{amssymb}
\usepackage{amsmath}
\usepackage{ifthen}
\usepackage{caption}
\usepackage[bookmarks=false]{hyperref}
\usepackage{url,moreverb,xspace}
\usepackage{tcolorbox}
\usepackage{enumitem}
\usepackage{array,graphicx}
\usepackage{soul}
\usepackage{balance}
\usepackage{pifont}
\usepackage{nicefrac}
\usepackage{mathtools}
\usepackage{tabularx}
\usepackage{xcolor,pifont}
\usepackage{tikz}
\usepackage{svg}
\usepackage{pdfpages}
\usepackage{array}
\usepackage{subcaption}
\usepackage{microtype}

\newcommand*\colourcheck[1]{%
  \expandafter\newcommand\csname #1check\endcsname{\textcolor{#1}{\ding{52}}}%
}
\newcommand*\colourcross[1]{%
  \expandafter\newcommand\csname #1cross\endcsname{\textcolor{#1}{\ding{56}}}%
}
\usepackage[vlined, boxruled, linesnumbered] {algorithm2e}
\SetKw{KwBy}{by}
\SetKw{KwBreak}{break}
\SetKw{KwReturn}{return}
\def\HiLi{\leavevmode\rlap{\hbox to \hsize{\color{gray!35}\leaders\hrule height .8\baselineskip depth .5ex\hfill}}}

\newcommand{\davetwo}{\mbox{DAVE-2}\xspace} %
\newcommand{\segformer}{\mbox{SegFormer}\xspace} %

\newcommand{\tool}{\mbox{PerturbationDrive}\xspace}

\newlength\BARWIDTH
\newlength\BARHEIGHT
\newcommand*\Chart[5]{%
  \begin{tikzpicture}[baseline={(current bounding box.south) - 10ex}] 
     Background with white fill
    \foreach \i [count=\xi] in {#1, #2, #3, #4, #5} {%
      \draw[fill=blue, draw=white, line width=0.1mm] (\xi*\BARWIDTH-\BARWIDTH, 0) rectangle (\xi*\BARWIDTH, \i*\BARHEIGHT); 
    }
    \draw[draw=black] (0, 0) rectangle (5*\BARWIDTH, 1.3*\BARHEIGHT); %
  \end{tikzpicture}%
}

\SetKwComment{Comment}{$\triangleright$\ }{}
\SetCommentSty{itshape}
\newboolean{showcomments}
\setboolean{showcomments}{false}
\ifthenelse{\boolean{showcomments}}
{\newcommand{\nb}[2] {
  \fcolorbox{black}{gray!20}{\bfseries\sffamily\scriptsize#1:}
  {\sf\small$\blacktriangleright$\textit{#2}$\blacktriangleleft$}
}
}
{\newcommand{\nb}[2]{}
}

\newcommand{\head}[1]{\noindent\textbf{#1.}}

\newcounter{fcounter}
\newcommand{\curl}[1]{\footnote{\url{#1}}}

\makeatletter
\newcommand{\thickhline}{%
    \noalign {\ifnum 0=`}\fi \hrule height 1pt
    \futurelet \reserved@a \@xhline
}
\makeatother

\begin{document}

\title{Benchmarking Image Perturbations for Testing Automated Driving Assistance Systems}

\author{\IEEEauthorblockN{Stefano Carlo Lambertenghi}
\IEEEauthorblockA{
\textit{Technical University of Munich, fortiss}\\
Munich, Germany \\
stefanocarlo.lambertenghi@tum.de, \\lambertenghi@fortiss.org}
\and
\IEEEauthorblockN{Hannes Leonhard}
\IEEEauthorblockA{
\textit{Technical University of Munich}\\
Munich, Germany \\
hannes.leonhard@tum.de}
\and
\IEEEauthorblockN{Andrea Stocco}
\IEEEauthorblockA{
\textit{Technical University of Munich, fortiss}\\
Munich, Germany \\
andrea.stocco@tum.de, \\stocco@fortiss.org}
}

\IEEEoverridecommandlockouts
\IEEEpubid{\makebox[\columnwidth]{XXX/\$31.00~\copyright2024 IEEE \hfill} \hspace{\columnsep}\makebox[\columnwidth]{ }}

\maketitle

\IEEEpubidadjcol

\begin{abstract}
Advanced Driver Assistance Systems (ADAS) based on deep neural networks (DNNs) are widely used in autonomous vehicles for critical perception tasks such as object detection, semantic segmentation, and lane recognition. However, these systems are highly sensitive to input variations, such as noise and changes in lighting, which can compromise their effectiveness and potentially lead to safety-critical failures. 

This study offers a comprehensive empirical evaluation of image perturbations, techniques commonly used to assess the robustness of DNNs, to validate and improve the robustness and generalization of ADAS perception systems. We first conducted a systematic review of the literature, identifying 38 categories of perturbations. Next, we evaluated their effectiveness in revealing failures in two different ADAS, both at the component and at the system level. Finally, we explored the use of perturbation-based data augmentation and continuous learning strategies to improve ADAS adaptation to new operational design domains. Our results demonstrate that all categories of image perturbations successfully expose robustness issues in ADAS and that the use of dataset augmentation and continuous learning significantly improves ADAS performance in novel, unseen environments.

\end{abstract}

\begin{IEEEkeywords}
autonomous driving systems testing, robustness
\end{IEEEkeywords}

\section{Introduction}\label{sec:introduction}

Advanced Driver Assistance Systems (ADAS) heavily rely on perception systems (e.g., cameras, LiDAR, and other sensors) to perceive complex, dynamic environments in real-time. 
These systems adopt deep neural networks (DNNs) for interpreting sensor data to assist tasks such as object detection, image classification, semantic segmentation, and regression, to enable accurate real-time driving functions~\cite{survey-lei-ma,yurtsever2020survey,li2024panopticperceptionautonomousdriving,grigorescu2020survey}. 

Despite their effectiveness in driving image understanding, these systems are expected to operate reliably across a large group of domain environments and operational domains. 
However, it is still infeasible to collect all representative scenarios during data collection and training campaigns. 
Thus, after deployment and in-field operation, the ADAS is likely to encounter inputs that significantly differ from the training data. Particularly, DNN-based ADAS perception systems are highly sensitive to input variations, such as lighting, environmental changes, noise, changes in lighting, or minor shifts in perspective~\cite{li2024panopticperceptionautonomousdriving}. These factors can lead to significant prediction errors~\cite{dodge2016understanding,geirhos2020generalisation}, misclassifications, or inaccurate segmentations. Such errors can propagate to the vehicle's decision-making modules, potentially resulting in safety-critical failures.

In literature, synthetic image perturbations have been proposed and utilised to assess and enhance the robustness of DNNs~\cite{hendrycks2019benchmarking,hendrycks2020augmix,rusak2020simple,Laermann-2019,8388338}. Image perturbations introduce controlled distortions to input images (e.g., by reducing the brightness or by blackening certain pixels), and have been used to simulate out-of-distribution conditions that challenge the robustness of the DNN to slight input variations~\cite{hendrycks2019benchmarking,hendrycks2020augmix,2023-Stocco-TSE,ayerdi2023metamorphic}. Additionally, image perturbations have been used for robustness/adversarial training, e.g., as a data augmentation strategy to enrich the training dataset with perturbed versions of the original, unperturbed images. This increases the diversity of the training set and helps the DNN to be invariant to various types of distortions, thereby improving its robustness.

In the context of ADAS testing, image perturbations have been employed by solutions such as DeepXplore~\cite{deepxplore}, DeepTest~\cite{deeptest} and DeepBillboard~\cite{deepbillboard}, using input transformations such as lighting changes and occlusions, real-world inspired synthetic perturbations like rain and fog, or synthetic adversarial billboards. These works target \textit{offline} ADAS testing of perception systems, which has been shown to be inadequate at revealing system-level failures~\cite{2020-Haq-ICST,briand-offline-emse,2023-Stocco-EMSE}.
Among the \textit{online} approaches for system-level ADAS testing, most research has focused on test generation to assess generalizability~\cite{NeelofarTOSEM,NeelofarICSE,pafot,epitester,lu2023deepqtesttestingautonomousdriving,pan2023safetyassessmentvehiclecharacteristics,KLUCK2023107225,10685189,HUMENIUK2023102990,10.1145/3650105.3652296,ARCAINI2024103171,10.1145/3550270,10.1007/978-3-031-49269-3_14,10.1007/978-3-031-49266-2_6,2025-Baresi-ICSE}, while fewer studies have addressed robustness testing at the system level. Some work proposed real-time adversarial attacks~\cite{2023-Stocco-TSE,dataAugment2020Liu,yoon2023learning}, state-aware billboards~\cite{DeepManeuver}, or generic sets of perturbations for runtime anomaly detection~\cite{ayerdi2023metamorphic,Luan2023Efficient}.
Although these solutions have shown great potential in exposing many failures, a comprehensive empirical evaluation of different image perturbations is still missing in the literature. Additionally, an in-depth analysis of their effectiveness, particularly in relation to their latency when used at a system level and their tuning in terms of intensity, remains an open area of investigation for efficient software testing.

To address this, in this paper, we provide the most comprehensive evaluation of image perturbations, benchmarking numerous perturbation techniques from the literature, across two popular ADAS tasks with increasing complexity. 
We first reviewed the literature on existing DNN image perturbations, identifying the most commonly used techniques for robustness evaluation. Then, we selected perturbations based on their feasibility and relevance to autonomous driving scenarios, retaining 32 types. To ensure a more thorough assessment, we evaluated the perturbations according to different intensity levels, enabling fine-grained control over their severity while ensuring that the original image semantics are preserved.

We systematically evaluate the performance of these perturbations in terms of their ability to generate robustness failure inputs in two ADAS, considering factors such as efficiency and effectiveness. Our findings confirm that all image perturbations expose ADAS failures, at different levels of severity, both at a component and a system level. 
Moreover, despite some of these image perturbations reflecting specific corner-case operational conditions (e.g., extreme lighting or camera occlusion), many do not represent naturalistic perturbations (i.e., weather effects) that can occur during real-world driving. Thus, we have investigated the usefulness of existing common image perturbations for domain adaptation and generalizability during robustness retraining.
Our findings reveal that adversarial retraining and continuous learning with common image perturbations allow ADAS to adapt to new environmental and naturalistic perturbations, thereby enhancing their robustness in real-world scenarios.

\noindent
This paper makes the following main contributions:
\begin{itemize}
    \item An empirical study comparing 32 image perturbations on two ADAS tasks, both at a component and at a system level. Our study identifies perturbation categories and intensities that are most effective for robustness failure exposure and retraining for domain adaptation.   
    \item A library for ADAS robustness testing, called \tool, which integrates several dozens of image perturbations for both offline and online robustness ADAS testing. \tool can be used as a standalone library for offline testing, providing perturbations that can be applied to input images. It also offers APIs that integrate seamlessly into different driving simulators without requiring modifications to existing infrastructures, and it is designed to be highly modular and extensible. To encourage open research, we made our library and experimental data available~\cite{tool}.
\end{itemize}

\section{DNN Image Perturbations}\label{sec:pert}

\subsection{Literature Review}\label{sec:literature}

We performed a systematic review of papers from Scopus and arXiv. Scopus was selected for its comprehensive archive of peer-reviewed articles, while arXiv allowed us to include grey literature and preprints in our review. These platforms helped us systematically search for papers that mentioned the keywords ``DNN'', ``image perturbation'' and ``ADAS'' within their abstract. 
This search found 2,943 papers, of which 1,814 in Scopus and 1,129 in arXiv.
The search results were each ranked based on the number of citations and the publication year, with more recent papers prioritized, producing a list of ranked studies, which we reviewed starting from the highest-ranked paper. For each source, we identified all proposed image perturbations and added them to a catalogue of perturbations. If the paper referenced any additional sources related to perturbations, these were also ranked similarly and added to the study list for further review. We continued this process, moving through the ranked studies until the catalogue of perturbations remained unchanged for ten consecutive studies. This search yielded 20 relevant sources, listed in \autoref{tab:grid_perturbation_categories_full}.


\begin{table*}[t!]

\caption{DNN image perturbation types identified in this study and their sources.}
\label{tab:grid_perturbation_categories_full}
\scriptsize
\centering
\begin{tabularx}{\textwidth}{XXXXXXXX}

\toprule


\textbf{Noise Perturbations (A)} & \textbf{Blur Perturbations (B)} & \textbf{Focus Perturbations (C)} & \textbf{Weather Perturbations (D)} & \textbf{Affine Transformations (E)} & \textbf{Graphic Patterns (F)} & \textbf{Color/Tone Adjustments(G)} & \textbf{Generative-based (H)}\\ 

\midrule

Gaussian Noise \newline  \cite{dodge2016understanding, geirhos2020generalisation, hendrycks2019benchmarking, rusak2020simple, Laermann-2019, ayerdi2023metamorphic,Luan2023Efficient,  michaelis2020benchmarking} \newline \scriptsize
Poisson Noise \newline  \cite{hendrycks2019benchmarking, rusak2020simple, ayerdi2023metamorphic, mu2019mnistc} \newline \scriptsize
Impulse Noise \newline  \cite{ hendrycks2019benchmarking, rusak2020simple, ayerdi2023metamorphic, michaelis2020benchmarking,mu2019mnistc} \newline \scriptsize
JPEG  \cite{dodge2016understanding, geirhos2020generalisation, hendrycks2019benchmarking, rusak2020simple, Laermann-2019,  ayerdi2023metamorphic, Luan2023Efficient, michaelis2020benchmarking} \newline \scriptsize
Speckle Noise \newline  \cite{rusak2020simple, ayerdi2023metamorphic} &

Defocus Blur \newline  \cite{ hendrycks2019benchmarking, rusak2020simple, ayerdi2023metamorphic,michaelis2020benchmarking} \newline \scriptsize
Motion Blur \newline  \cite{ hendrycks2019benchmarking, rusak2020simple, Luan2023Efficient, mu2019mnistc,michaelis2020benchmarking} \newline \scriptsize
Zoom Blur \newline  \cite{hendrycks2019benchmarking, rusak2020simple,  Luan2023Efficient,michaelis2020benchmarking} \newline \scriptsize
Gaussian Blur  \cite{ayerdi2023metamorphic,deeptest,Luan2023Efficient} \newline \scriptsize
Low Pass Filter~\cite{geirhos2020generalisation} \scriptsize&

Frosted Glass  \cite{ hendrycks2019benchmarking, rusak2020simple, Luan2023Efficient,michaelis2020benchmarking} \newline \scriptsize
Snow  \cite{ hendrycks2019benchmarking, Luan2023Efficient,michaelis2020benchmarking} \newline \scriptsize
Fog  \cite{ hendrycks2019benchmarking, rusak2020simple, ayerdi2023metamorphic,Luan2023Efficient, michaelis2020benchmarking} \newline \scriptsize
Brightness  \cite{hendrycks2019benchmarking,ayerdi2023metamorphic, deepxplore,deeptest,michaelis2020benchmarking,mu2019mnistc,  8953317, cubuk2019randaugment} \newline \scriptsize
Contrast  \cite{dodge2016understanding, geirhos2020generalisation, hendrycks2019benchmarking, rusak2020simple, 8388338,ayerdi2023metamorphic,deeptest, michaelis2020benchmarking, 8953317, cubuk2019randaugment} \scriptsize &

Elastic  \cite{ hendrycks2019benchmarking, rusak2020simple,michaelis2020benchmarking} \newline \scriptsize
Pixelate  \cite{ hendrycks2019benchmarking, rusak2020simple,michaelis2020benchmarking} \newline \scriptsize
Sample~Pairing \cite{hendrycks2020augmix,ayerdi2023metamorphic, 8953317} \newline \scriptsize
Sharpen  \cite{8388338,8953317, cubuk2019randaugment} \scriptsize &

Shear Mapping \newline  \cite{hendrycks2020augmix, 8388338,deeptest, mu2019mnistc, 8953317, cubuk2019randaugment} \newline \scriptsize
Scale  \cite{ Laermann-2019, 8388338,deeptest} \newline \scriptsize
Translate  \cite{hendrycks2020augmix,Laermann-2019,deeptest, 8953317, cubuk2019randaugment} \newline \scriptsize
Rotate  \cite{geirhos2020generalisation, hendrycks2020augmix, Laermann-2019,8388338,deeptest, 8953317, cubuk2019randaugment} \newline \scriptsize
Reflection  \cite{8388338} \scriptsize &

Splatter  \cite{ayerdi2023metamorphic,mu2019mnistc} \newline \scriptsize
Dotted Lines  \cite{mu2019mnistc} \newline \scriptsize
ZigZag  \cite{mu2019mnistc} \newline \scriptsize
Canny Edges  \cite{mu2019mnistc} \newline \scriptsize
Cutout  \cite{hendrycks2020augmix, deepxplore,8953317} \scriptsize &

False color  \cite{geirhos2020generalisation,cubuk2019randaugment} \newline \scriptsize
Phase~scrambling~\cite{geirhos2020generalisation} \newline \scriptsize
Histogram equalization \cite{geirhos2020generalisation,8388338,8953317,cubuk2019randaugment} \newline \scriptsize
White balance  \cite{8388338} \newline \scriptsize
Greyscale  \cite{geirhos2018biasedTexture} \newline \scriptsize
Saturation  \cite{ayerdi2023metamorphic,michaelis2020benchmarking} \newline \scriptsize
Posterize  \cite{hendrycks2020augmix,8953317} \scriptsize
&

Cycle-consistent   \cite{NEURIPS2019-b83aac23,8863362,10.1145/3238147.3238187} \newline \scriptsize
Style-transfer \newline \cite{dataAugment2020Liu,michaelis2020benchmarking,geirhos2018biasedTexture} \scriptsize\\

\bottomrule
\end{tabularx}
\end{table*}

\subsection{List of DNN Image Perturbations}\label{sec:list}

\autoref{tab:grid_perturbation_categories_full} presents the list of 38 perturbations obtained with our review of the literature, and their sources. 
We evaluated perturbations based on visual similarities or computational techniques and assigned them to eight main categories: noise, blur, focus, weather-related changes, affine transformations, graphic patterns, colour and tone adjustments, and generative-based. In the following, we briefly describe each category.

\subsubsection{Noise Perturbations (A)}

This category includes random variations in pixel values or image graininess, typically resulting from electronic noise. This category includes: (A-I-II)~Gaussian noise and Poisson noise consist in the addition of statistical noise to an image, using the probability density functions of the Normal distribution or the Poisson distribution, respectively. 
(A-III)~Impulse noise consists of random, sharp and sudden disturbances, taking the form of scattered bright or dark pixels. (A-IV)~JPEG artifacts perturbs an image in the same manner as JPEG compression artifacts would. (A-V)~Speckle Noise adds granular noise textures to the image.

\subsubsection{Blur and Focus Perturbations (B)}

These perturbations cause a reduction in image sharpness or clarity, often stemming from improper camera settings. A key characteristic of these perturbations is the use of kernels, which calculate the perturbation by averaging or smoothing pixel values in specific areas of the image. Adjustments to the size of the kernel matrix or changes to its values can modify the intensity of these perturbations. We have identified the following types of perturbations:
(B-I)~Defocus blur simulates the effects of the camera lens being out of focus via disc-shaped kernels.
(B-II)~Motion blur mimics the streaking effects caused by moving objects.
(B-III)~Zoom blur simulates a radial blur that emanates from a central point of the image.
(B-IV)~Gaussian blur blurs the image by applying the Gaussian function on the image.
(B-V)~Low-pass filter calculates the average of each pixel to its neighbours.

\subsubsection{Weather Perturbations (C)}

These perturbations simulate weather conditions such as snow, rain, or fog. Such conditions are influenced by specific times of the day or seasons; for instance, images taken at sunrise or sunset might exhibit enhanced brightness or contrast, while winter could naturally bring about snowfall.
This category includes:
(C-I)~Frosted glass simulates the effects of frosting on a camera lens.
(C-II)~Snow simulates the presence of snow crystals.
(C-III)~Fog reduces the image's contrast and saturation to simulate the presence of fog.
(C-IV)~Brightness changes the image's brightness intensity, to simulate changes of environment lighting.
(C-V)~Contrast increases the difference of luminance on the image.

\subsubsection{Distortion Perturbations (D)}

Distortion perturbations randomly displace or overlap image pixels, leading to distorted figures and shapes in the visuals. 
We identify the following perturbations: (D-I)~Elastic moves each image's pixel by a random offset derived from a Gaussian distribution. (D-II)~Pixelate divides the image in a set of pixel regions and sets the average pixel value of the region to all pixels in it. (D-III)~Sample pairing randomly samples two regions of the image and blends them together with a varying alpha value. (D-IV)~Sharpen removes blurring by using the sharpen kernel.

\subsubsection{Affine Transformations (E)} 

These transformations preserve the collinearity and parallelism of lines within the image, meaning that straight lines remain straight, and sets of parallel lines remain parallel after the transformation. However, the actual distances between points and the angles between lines can change. This allows for transformations such as rotation, scaling, translation, and shearing, fundamentally altering the image’s appearance while maintaining a level of geometric consistency. These include: (E-I)~Shear mapping shifts each point in an image horizontally. The shift's direction and magnitude are based on each point's perpendicular distance from a reference line parallel to the shift direction, resulting in a slanted or skewed appearance of the image. (E-II)~Scale increases or decreases the size of the image by a certain factor. (E-III)~Translate moves all the pixels of an image in a certain direction. (E-IV)~Rotate rotates the image by a certain angle in the Euclidean space. (E-V)~Reflection creates a mirror effect by appending a flipped version of the image at a certain height.

\subsubsection{Graphic Patterns (F)} 

This type of perturbation involves inserting repetitive graphics and patterns randomly across the image. A pattern in this context is defined as a specific shape, like a dot or a rectangle, that is systematically repeated throughout the image. This category includes: (F-I)~Splatter randomly adds black patches of varying size to the image. (F-II)~Dotted lines randomly adds straight dotted lines to the image. (F-III)~ZigZag randomly adds black zig-zagging lines to the image. (F-IV)~Canny edge filter applies canny edge detection to highlight images and lay them over the images. (F-V)~Cutout inserts black rectangular shapes on the image.

\subsubsection{Color and Tone Adjustments (G)} 

These perturbations modify the color and tone characteristics of the image by averaging, increasing, or decreasing specific channels.
We have identified the following perturbations: (G-I)~False color filter swaps color channels, inverts color channels, or averages color channels with each other. (G-II)~Phase scrambling  scrambles image channels using the Fast Fourier Transform. (G-III)~Histogram equalization enhances the image contrast by spreading the pixel intensities using the image histogram. (G-IV)~White balance globally adjusts the intensity of colors to adjust white portions of the image. (G-V)~Greyscale filter converts the image to greyscale. (G-VI)~Saturation increases or decreases the saturation of the image by changing the S channel of the image in HSV (Hue, Saturation, Lightness) representation. (G-VII)~Posterize reduces the number of distinct colors by quantizing the color channels.

\begin{figure*}[t]
  \centering
  \begin{subfigure}{0.48\textwidth}
    \includegraphics[width=\linewidth]{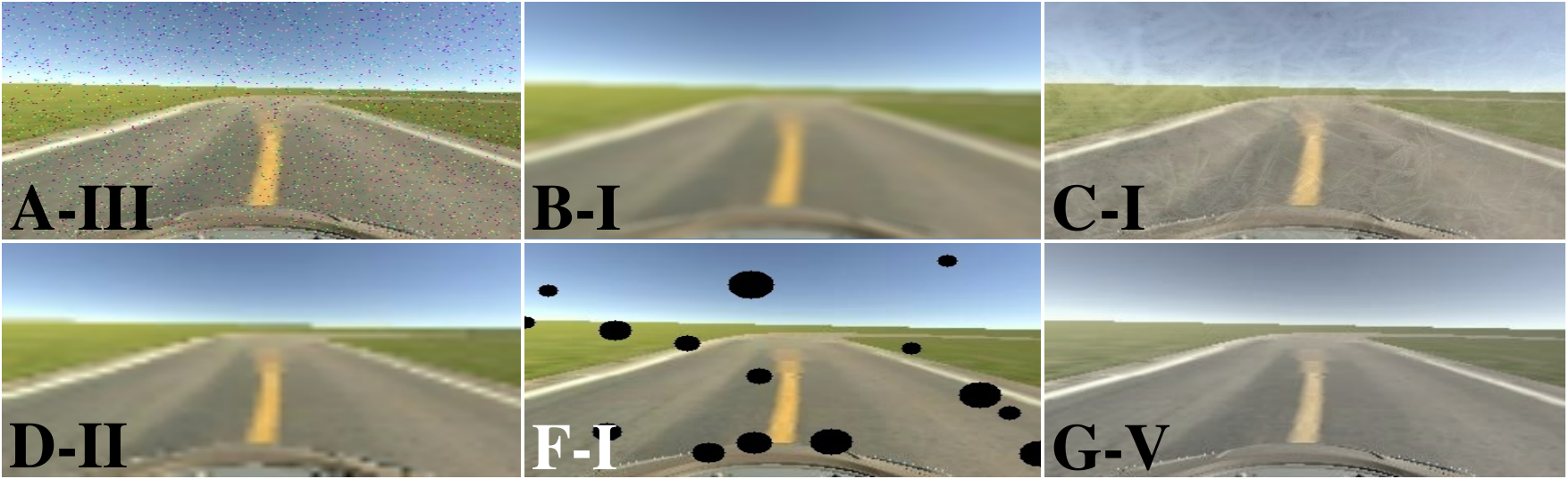}
    \caption{Valid.}
  \end{subfigure}
  \hspace{0.01\textwidth}
  \begin{subfigure}{0.48\textwidth}
    \includegraphics[width=\linewidth]{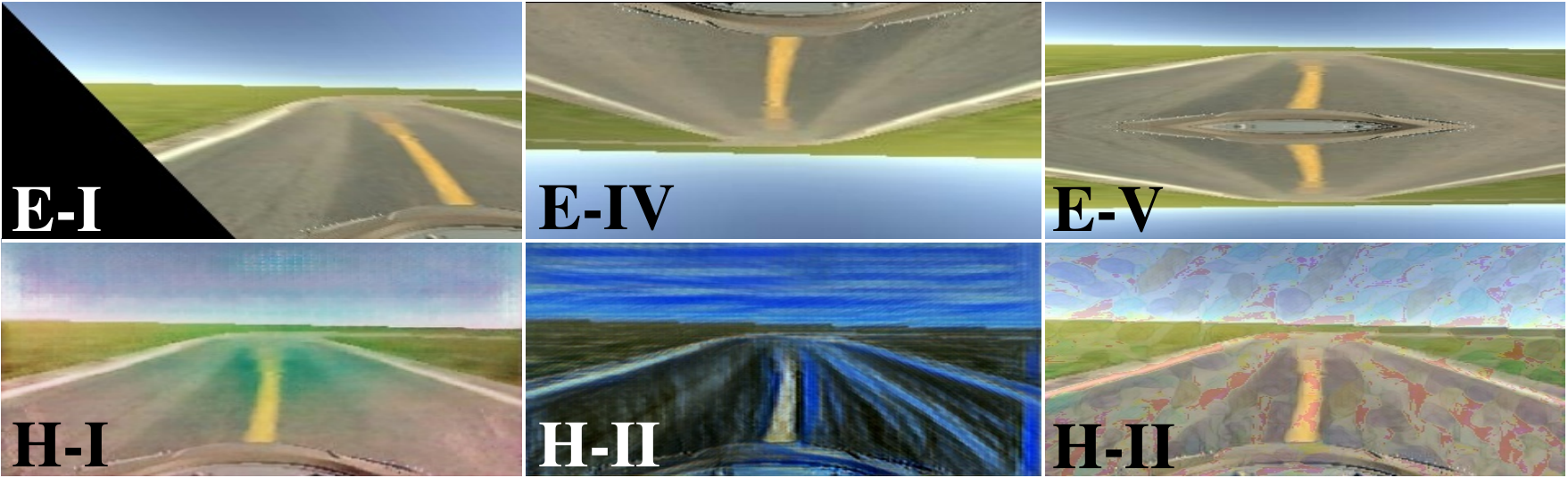}
    \caption{Invalid.}
  \end{subfigure}
  \caption{Valid and invalid perturbation types.}
  \label{fig:invalid-perturbations}
\end{figure*}

\subsubsection{Generative-based Perturbations (H)}

The perturbations discussed in this section use generative models to modify images, with the goal of changing the look of the input domain into the look of another alternative domain. In this category, we consider: (H-I)~cycle-consistent models enable the generation of images across two image distributions representing different domains. (H-II)~Style-transfer models apply artistic styles to images, using pre-trained generative models.

\subsection{Image Perturbations Validity}

We manually analyzed each perturbation type with the aim to keep those that generate images that maintain the semantic between the original and the augmented image and that represent valid driving images (e.g., \autoref{fig:invalid-perturbations}(a)). We implemented these checks to make sure to use perturbations that do not drastically change the image content (i.e., making the road no longer visible, or an image in which the road is flipped vertically, see \autoref{fig:invalid-perturbations}(b)~E-IV), thereby exposing failures that are not relevant. 

In particular, we filtered out perturbations such as some types of affine transformations, rotations, shear mapping, reflection, and generative based. Affine transformations can distort the image to the extent that generating a valid driving command, even for a human, becomes impossible. For example, shear mapping (E-I) skews and distorts the image, pushing parts outside the frame and potentially altering the ADAS's driving decisions. Rotations (E-IV) can shift images by an angle $\theta$, with larger angles (e.g., $\theta=180^\circ$) inverting the image, while reflections (E-V) duplicate content, both of which may confuse DNNs and mislead the ADAS. Similarly, generative-based perturbations (H) were excluded. CycleGAN (H-I) relies on the input domain matching the source domain of its training dataset, making it unsuitable for general driving scenarios, while style-transfer (H-II) often introduces unrealistic alterations, such as exaggerated textures and colour shifts, distorting the visual content to an unrealistic degree (\autoref{fig:invalid-perturbations}).
For the segmentation task, we excluded the entire Affine Transformations (E) group because validating the model’s performance would require distorting the pixel-level ground truth classes to match the transformations, which is not feasible for accurate evaluation.

To obtain the five intensity levels for the included perturbations, we resort to a visual assessment. Specifically, we incrementally applied each perturbation until we could no longer understand the depicted scene. The intensity level right before this threshold was considered as the maximum intensity and we discretized the intensity range into five uniform steps. 

\subsection{Implementation}\label{sec:approach}

To support our experimental evaluation, we develop an extensible Python library called \tool, which is available~\cite{tool}. It systematically enables the application of a large variety of image perturbations for conducting both model-level and system-level robustness testing of ADAS. 
Architecturally, \tool consists of three main components: an image perturbation module, a simulator interface (only for the case of system-level testing), and a benchmarking controller. The image perturbation component implements all perturbations described in \autoref{sec:pert} and applies them to a given input image. It also allows for the integration of new perturbations through the extension of an abstract interface.

For online testing, the framework integrates seamlessly with driving simulators. The initial release of \tool supports ADAS models developed in TensorFlow/Keras and is compatible with two Unity-based driving simulators: the Udacity Simulator~\cite{udacity-simulator} and the Sdsandbox Donkey Car\textsuperscript{\texttrademark} simulator~\cite{sdsandbox}. It enables the generation of different road layouts, represented as a series of waypoints~\cite{2025-Ali-ICSEW}, and allows the application of perturbations in real time.

Finally, the benchmarking controller manages the testing process. For offline testing, it applies perturbations to ADAS input images and compares the ADAS responses with either ground-truth values or its output on unperturbed images. During online testing, the framework logs various metrics such as the ADAS actions, perturbed and unperturbed images, and vehicle speed to determine if the system successfully drives the scenario or encounters a failure.

\section{Empirical Study}\label{sec:empirical-study}

\subsection{Research Questions}\label{sec:rqs}

\noindent
\textbf{RQ\textsubscript{1} (effectiveness):} \textit{Which types of image perturbations are more effective in inducing robustness failures in ADAS?}

The first research question investigates how different types of image perturbations, applied at varying levels of intensity, impact the reliability of ADAS. 

\noindent
\textbf{RQ\textsubscript{2} (generalization):} \textit{How effective are common image perturbations in enhancing the generalization of ADAS to more naturalistic perturbations?}

The second research question investigates how common perturbations can improve the generalization ability of ADAS to real-world environmental scenario changes, i.e., weather conditions, through retraining. 

\subsection{Objects of Study}

We consider NHTSA~\cite{nhtsa} Level 2 ADAS that perform vision-based perception tasks, i.e., from data gathered by camera sensors of a vehicle. Despite the adoption of Level 2 ADAS in many commercial vehicles, their reliability remains a concern, as evidenced by numerous recent crash reports~\cite{NHTSA-level2-crashes}. 
We focus on two specific ADAS applications: a system for semantic segmentation and another designed for lane-keeping and adaptive cruise control (LK/ACC).

\subsubsection{Semantic Segmentation}

\segformer~\cite{segformer} is a vision transformer-based model designed for semantic segmentation, where each pixel in an input image is classified into one of several object classes. The model employs a hierarchical transformer architecture for feature extraction and a multi-level feature aggregation network to generate segmentation maps with both fine detail and global context. Unlike traditional convolutional models, \segformer omits positional encodings, enhancing its efficiency and scalability for real-time applications, including ADAS. Trained on large-scale datasets like Cityscapes~\cite{cityscapes}, \segformer has demonstrated competitive performance in segmenting complex driving environments~\cite{2024-Lambertenghi-ICST}.

\subsubsection{LK/ACC}

\davetwo is a convolutional neural network developed for multi-output regression tasks based on imitation learning~\cite{nvidia-dave2}. The model architecture includes three convolutional layers for feature extraction, followed by five fully connected layers. 
\davetwo has been extensively used in a variety of ADAS testing studies~\cite{2023-Stocco-TSE,deeptest,10.1145/3238147.3238187,biagiola2023better,2021-Jahangirova-ICST,biagiola2023boundary}.
The model takes as input an image representing a road scene, and it is trained to predict vehicle's actuators commands. Our implementation includes a DNN with lane-keeping (LK) and adaptive cruise control (ACC) capabilities, as \davetwo is trained to conduct the vehicle on the right lane of the road at the maximum possible speed, by predicting appropriate steering and throttle commands. 

\subsection{Experimental Platforms and Benchmarks}

\subsubsection{Semantic Segmentation}\label{sec:dataset}

We test SegFormer using the Virtual KITTI dataset (vKITTI)~\cite{vkitti}, commonly used for autonomous driving research. It provides 21,260 photo-realistic frames across five of the 20 KITTI real-world scenarios (i.e. 01, 02, 06, 18, and 20) rendered using the Unity engine. It includes pixel-level segmentation ground truths and semantic labels for urban objects such as roads, traffic lights and vehicles. Each frame is available in six weather conditions: sunny (nominal), fog, morning, overcast, rain, and sunset. We divided the scenarios in vKITTI as follows:

\begin{itemize}
    \item \textit{Training set}: Scenarios 01, 02, and 06 (nominal weather), randomly split into $90\%$ and $10\%$ of the samples for training and validation of the \segformer model for \textbf{RQ\textsubscript{1}};
    \item \textit{Augmentation set}: Scenario 18 (nominal weather), split similarly, used for fine-tuning \segformer with perturbations for \textbf{RQ\textsubscript{2}};
    \item \textit{Testing set (N)}: Scenario 20 (nominal weather), used to evaluate perturbation disruptions for \textbf{RQ\textsubscript{1}} and as a baseline for \segformer variants in \textbf{RQ\textsubscript{2}};
    \item \textit{Testing set (W)}: Scenario 20 (with weather effects), used to evaluate \segformer variants for \textbf{RQ\textsubscript{2}}.
\end{itemize}

\subsubsection{LK/ACC}\label{sec:sims}

To evaluate the generalizability of our results, we conducted experiments on both the Udacity~\cite{udacity-simulator} and Donkey Car\textsuperscript{\texttrademark}~\cite{sdsandbox} simulation environments. 

Udacity~\cite{udacity-simulator} is developed with Unity 3D~\cite{unity}, a popular cross-platform game engine, based on the Nvidia PhysX engine~\cite{PhysX}, featuring discrete and continuous collision detection, ray-casting, and rigid-body dynamics simulation. 
Udacity also supports testing under various weather conditions, such as day/night, rain, snow, and fog, which we refer to as \textit{naturalistic} perturbations because they simulate real-world phenomena like virtual artefacts (e.g., raindrops) and lighting variations (e.g., day/night transitions). In addition to these effects, in this study, we introduce four new weather conditions that are not based on particle effects but instead focus on lighting changes by altering the skybox and scene illumination intensity. These conditions, listed in order of increasing darkness, are named: dawn, moonshine, starry, and dark/overcast.
Donkey Car\textsuperscript{\texttrademark} includes a high fidelity digital twin of the Donkey Car, a 1:16 scale radio-controlled car with self-driving capabilities, used for ADAS testing research in physical environments~\cite{2023-Stocco-TSE,2023-Stocco-EMSE}. 
We selected these platforms because they are open-source and suitable for Level~2 ADAS evaluation. However, this choice is not exclusive, and other simulators can be integrated in \tool with additional engineering cost.

For both simulators, we use the following scenarios:

\begin{itemize}
    \item \textit{Training roads:} These roads are used to train the \davetwo model evaluated in both \textbf{RQ\textsubscript{1}} and \textbf{RQ\textsubscript{2}}. The road structures are randomly generated using \tool, incorporating a variety of curves, lengths, and curvatures to ensure diversity in the training set.
    \item \textit{Testing roads\textsubscript{RQ1}:} These manually designed scenarios consist of 10 road tracks with increasing difficulty, ranging from simple, straight roads to more complex, curvy paths, used to evaluate the model's performance in \textbf{RQ\textsubscript{1}}.
\end{itemize}

As we evaluate generalizability (\textbf{RQ\textsubscript{2}}) using only Udacity, since the Donkey Car simulator lacks weather variations, we include additional independent test sets: 

\begin{itemize}
    \item\textit{Fine-tuning roads:} These roads are used to fine-tune the \davetwo model evaluated in \textbf{RQ\textsubscript{2}}, as such they are designed to differ from roads used to test the model's performance. We use \tool to generate random roads with specific curvature ranges and road lengths which differ from the ones found in other road sets. 
    \item\textit{Testing roads\textsubscript{RQ2}:} These roads are used to evaluate both the nominal and fine-tuned \davetwo models in \textbf{RQ\textsubscript{2}}. These scenarios include 15 road shapes, with different degrees of complexity, and the ability to set simulation-based weather effects. In particular, five of these scenarios drastically differ from the ones found in \textit{Training roads}.
\end{itemize}

\subsection{Procedure}

As one of the goals of our study is to evaluate the image perturbations for system level testing, we performed a preliminary analysis to assess their computational overhead. Excessive processing times could in fact jeopardize the execution of simulation-based tests and lead to spurious failures that are not related to the actual robustness of the ADAS. 
We evaluated the execution time of 250 iterations for each type of perturbation across multiple intensity levels, using random RGB images with a resolution of 240$\times$320 pixels—consistent with the image dimensions used by the simulators under evaluation. This experiment was conducted using the \texttt{pyperf} library~\cite{pyperf} on a machine equipped with an Apple M1 processor. Since each perturbation must be applied to every simulation frame, we established an upper time limit for acceptable computation based on the simulators’ frame rates.

The two simulators in our study, Udacity and Sdsandbox Donkey Car, operate at frame rates of 20 and 30 frames per second (fps), respectively, which correspond to frame intervals of 50 ms and 33.3 ms. To ensure consistency in comparisons, we adopted the higher frame rate of 30 fps, setting 33.3 ms as the maximum acceptable computation time per frame for all perturbations. This upper limit assumes that the time required for the ADAS system to process each frame is negligible.

\begin{figure}[t!]
  \centering
  \includegraphics[width=1.0\linewidth]{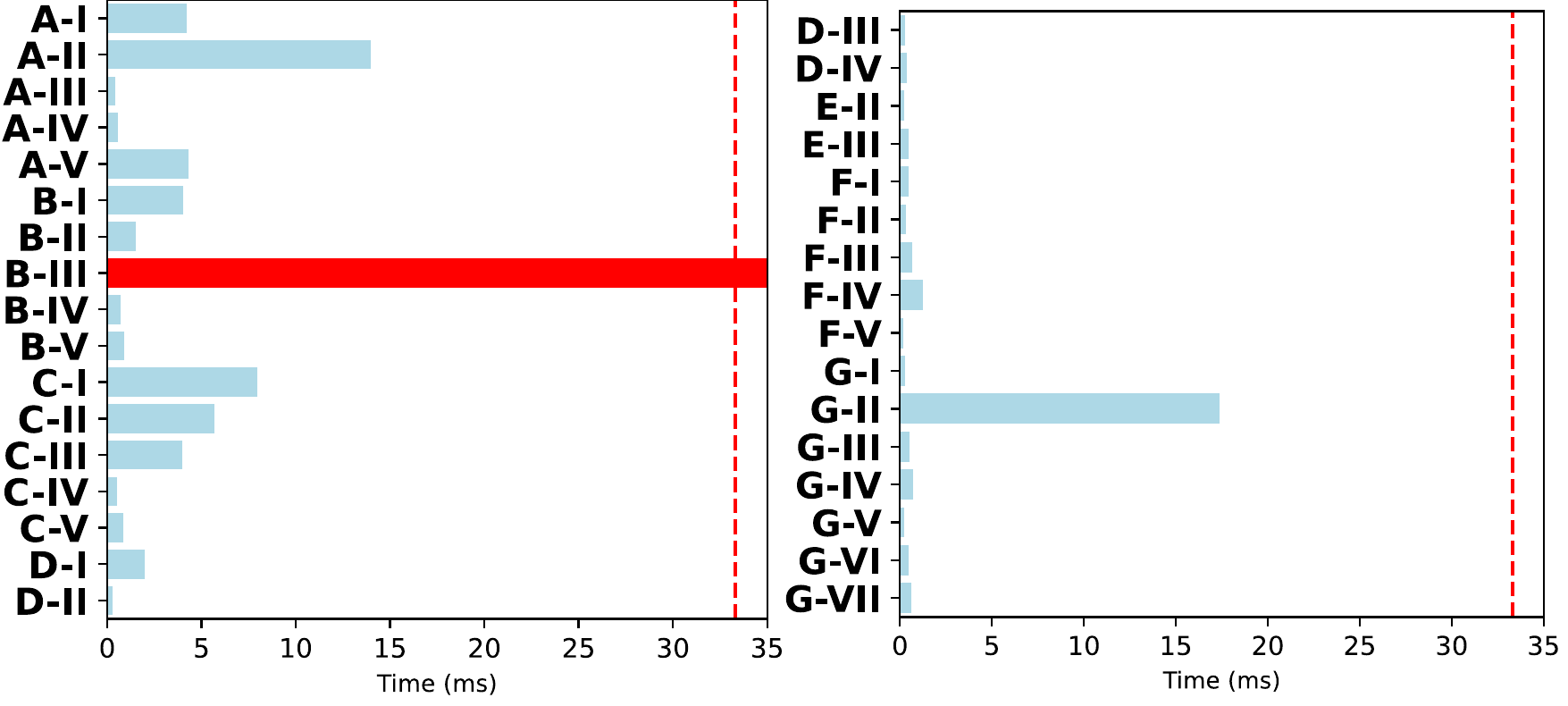}
  \caption{Benchmarking perturbations.}
  \label{fig:benchmark}
\end{figure}

\autoref{fig:benchmark} reports the average execution time for each perturbation. Our study shows that the majority of perturbations are feasible for real-time evaluation, with most taking less than 10 ms to execute. Only Zoom blur (B-III) significantly exceeds the 33.3 ms threshold at 95.6 ms. As a result, Zoom blur is excluded from further evaluations.

\subsubsection{RQ\textsubscript{1}}

\head{Semantic Segmentation}
We first fine-tune a pre-trained SegFormer model~\cite{a2017_nvidiasegformerb0finetunedcityscapes6401280} for 10 epochs using the \textit{training set} split of the vKITTI dataset (see \autoref{sec:dataset}) and the Adam optimizer with a learning rate of $6e^{-5}$. 

Regarding the evaluation phase, we instructed \tool to introduce controlled perturbations in the images of the \textit{Testing set (N)}. 
Each image was perturbed across five different intensity levels. 
This approach enables us to systematically evaluate not only how different perturbation types affect the model's output but also at which intensity levels the impact becomes significant.
We then execute the trained SegFormer model on both the nominal (unperturbed) and perturbed images of \textit{Testing set (N)}. For each test image, the model generates a segmentation map, with each pixel classified into a corresponding semantic category. These predicted segmentation maps are then compared with the ground truth annotations from the dataset to assess the effect of the perturbations on the model's segmentation performance.

To quantify the effects of the perturbations, we calculate 
the Intersection over Union (IoU), 
chosen for its widespread use in evaluating segmentation model performance, particularly for driving scenes~\cite{cityscapes-benchmark,kitti-bench}.
IoU measures, for each semantic class, the overlap between the predicted and ground truth segmentation maps, calculated as the ratio of the intersection (where predicted and true segments match) to the union (the total area covered by both). This makes IoU especially useful for understanding how accurately the model identifies critical elements in driving environments, such as vehicles. 

\head{LK/ACC}
We trained the \davetwo model in both simulators using data collected by a human driver on \textit{Training roads}. We then validate the model using the \textit{Testing roads\textsubscript{RQ1}}. The behaviour of the ADAS under nominal conditions constitutes a baseline for evaluating the effects of perturbations.

Next, we use \tool to inject perturbations into the images during simulation-based testing, on the same set of \textit{Testing roads\textsubscript{RQ1}}.
The \davetwo model's performance on each perturbed road is then compared against the performance under nominal conditions. 
To quantify the performance degradation caused by perturbations, we evaluate several metrics. First, we quantified the \textit{success rate} and \textit{completion rate}. The first metric indicates the percentage of scenarios in which the LK-ACC system successfully reaches the end goal, while the second measures the extent of the scenario executed before either a failure or goal completion occurs. Next, we classify failures into two types: \textit{Out of road (OR)}, when the vehicle leaves the designated driving lane, and \textit{Out of time (OT)} triggered when misbehaviour causes a delay, leading to a 200-second timeout before the scenario is completed.
For successful scenarios, we further analyze the \textit{execution time}, which is useful for identifying the quality of throttle predictions and the \textit{driving jitter}, which helps assess the steering quality by calculating the first derivative of the distance from the center of the lane (Cross-Track Error) and normalizing it by the lane width to indicate deviations in lane centering.


\begin{table*}[t]
\centering
\caption{RQ\textsubscript{1}: Effectiveness results for different types of perturbations for robustness testing of ADAS.}
\label{tab:RQ1}
\resizebox{\textwidth}{!}{ %

\begin{tabular}{lrrrrrrrrrrrrrrrrrr}

\toprule

& \multicolumn{4}{c}{Semantic Segmentation} 
& \multicolumn{14}{c}{LK/ACC} \\

\cmidrule(r){2-5}
\cmidrule(r){6-19}

& \multicolumn{4}{c}{vKITTI} 
& \multicolumn{7}{c}{Udacity} 
& \multicolumn{7}{c}{Donkey Car} \\

\cmidrule(r){2-5}
\cmidrule(r){6-12}
\cmidrule(r){13-19}

& &&&& \multicolumn{5}{c}{Overall} & \multicolumn{2}{c}{Non-failing} & \multicolumn{5}{c}{Overall} & \multicolumn{2}{c}{Non-failing}\\

\cmidrule(r){6-10}
\cmidrule(r){11-12}
\cmidrule(r){13-17}
\cmidrule(r){18-19}

& \multicolumn{4}{c}{IoU} 
& \multicolumn{3}{c}{Success Rate} 
& \multicolumn{2}{c}{Fail Type} 
& Time
& Jitter
& \multicolumn{3}{c}{Success Rate} 
& \multicolumn{2}{c}{Fail Type} 
& Time 
& Jitter \\

\cmidrule(r){2-5}
\cmidrule(r){6-8}
\cmidrule(r){9-10}
\cmidrule(r){11-12}
\cmidrule(r){11-12}
\cmidrule(r){13-15}
\cmidrule(r){16-17}
\cmidrule(r){18-19}

Perturbation & Avg & Std & Max & Min &
Avg & Std & Trend
& OR & OT 
& Avg  
& Avg  
& Avg & Std & Trend
& OR & OT 
& Avg  
& Avg  \\

\midrule

nominal & 0.66 & - & - & - & 100\% & - & - & 0 & 0 & 96.84 &  2.83\%   & 100\% & - & - & 0 & 0 & 102.17  & 2.50\% \\

\midrule

\textbf{Noise} & & & & & & & & & & & & & & & & & &  \\ 

\quad A-I Gaussian noise & 0.43 & 0.16 & 0.64 & 0.23                                          & 86\% & 3.00 & \Chart{0.90}{0.90}{0.90}{0.80}{0.80} & 2 & 5 & 95.6 & 44.8\%                   & 98\% & 2.00 & \Chart{1.00}{1.00}{1.00}{1.00}{0.90} &0 & 1 & 83.3 &  9.4\%\\            
\quad A-II Poisson noise& 0.50 & 0.10 & 0.58 & 0.32                                           & 100\% & 0.00 & \Chart{1.00}{1.00}{1.00}{1.00}{1.00} & 0 & 0 & 50.2  & 9.4\%                  & 98\% & 2.00 & \Chart{1.00}{1.00}{1.00}{1.00}{0.90} &1 & 0 & 52.2 & 9.0\% \\                 
\quad A-III Impulse noise& 0.41 & 0.16 & 0.62 & 0.21                                          & 96\% & 3.00 & \Chart{1.00}{1.00}{0.90}{0.90}{1.00} & 0 & 2 & 129.9 &  11.0\%                 & 100\% & 0.00 & \Chart{1.00}{1.00}{1.00}{1.00}{1.00} &0 & 0 & 95.5 &  7.6\% \\              
\quad A-IV JPEG artifacts& 0.59 & 0.02 & 0.62 & 0.56                                          & 66\% & 88.00 & \Chart{1.00}{0.80}{0.80}{0.40}{0.30} & 1 & 16 & 149.2 &  38.8\%             & 98\% & 2.00 & \Chart{0.90}{1.00}{1.00}{1.00}{1.00} &1 & 0 & 77.3 &  4.8\% \\              
\quad A-V Speckle noise& 0.58 & 0.09 & 0.66 & 0.41                                            & 88\% & 2.00 & \Chart{0.90}{0.90}{0.90}{0.90}{0.80} & 6 & 0 & 59.2  & 20.0\%                   & 100\% & 0.00 & \Chart{1.00}{1.00}{1.00}{1.00}{1.00} &0 & 0 & 73.4 &  6.2\% \\ [0.5em]                
                         
\textbf{Blur and Focus} & & & & & & & & & & & & & & & & & &  \\                         
                         
\quad B-I Defocus blur& 0.57 & 0.06 & 0.65 & 0.47                                             & 96\% & 3.00 & \Chart{1.00}{0.90}{1.00}{1.00}{0.90} & 1 & 1 & 103.4 &  12.6\%                         & 100\% & 0.00 & \Chart{1.00}{1.00}{1.00}{1.00}{1.00} &0 & 0 & 84.6 & 8.0\% \\               
\quad B-II Motion blur& 0.63 & 0.03 & 0.66 & 0.58                                             & 84\% & 8.00 & \Chart{1.00}{0.80}{0.80}{0.80}{0.80} & 0 & 8 & 122.2 &  46.0\%                        & 94\% & 8.00 & \Chart{1.00}{0.80}{0.90}{1.00}{1.00} &3 & 0 & 93.8 &  8.6\% \\                 
\quad B-IV Gaussian blur& 0.58 & 0.08 & 0.66 & 0.46                                           & 94\% & 3.00 & \Chart{1.00}{0.90}{0.90}{0.90}{1.00} & 3 & 0 & 87.6 &  11.8\%                         & 94\% & 3.00 & \Chart{1.00}{0.90}{0.90}{0.90}{1.00} &3 & 0 & 99.9 & 9.0\% \\               
\quad B-V Low-pass filter& 0.63 & 0.01 & 0.64 & 0.62                                          & 88\% & 2.00 & \Chart{0.80}{0.90}{0.90}{0.90}{0.90} & 6 & 0 & 88.0 &  28.6\%                 & 96\% & 3.00 & \Chart{1.00}{1.00}{1.00}{0.90}{0.90} &2 & 0 & 92.7 & 8.6\% \\ [0.5em]              
                                    
\textbf{Weather} & & & & & & & & & & & & & & & & & &   \\

\quad C-I Frosted glass& 0.62 & 0.04 & 0.65 & 0.54                                                   & 70\% & 135.00 & \Chart{1.00}{1.00}{0.70}{0.70}{0.10} & 15 & 0 & 52.1 &  11.6\%                           & 94\% & 3.00 & \Chart{0.90}{0.90}{0.90}{1.00}{1.00} &3 & 0 & 69.3  & 7.8\% \\        
\quad C-II Snow& 0.64 & 0.04 & 0.67 & 0.56                                                  & 30\% & 120.00 & \Chart{0.80}{0.50}{0.20}{0.00}{0.00} & 35 & 0 & 32.5 &  13.8\%                                   & 96\% & 8.00 & \Chart{1.00}{1.00}{1.00}{1.00}{0.80} &2 & 0 & 74.9  & 10.2\% \\               
\quad C-III Fog& 0.56 & 0.14 & 0.67 & 0.29                                                 & 54\% & 228.00 & \Chart{1.00}{1.00}{0.60}{0.10}{0.00} & 23 & 0 & 52.2  & 10.6\%                                 & 90\% & 5.00 & \Chart{0.90}{0.90}{1.00}{0.90}{0.80} &5 & 0 & 81.7  & 11.6\% \\                
\quad C-IV Brightess& 0.66 & 0.00 & 0.66 & 0.65                                                  & 92\% & 7.00 & \Chart{1.00}{1.00}{0.90}{0.90}{0.80} & 3 & 1 & 123.2 &  12.2\%                                & 96\% & 8.00 & \Chart{1.00}{1.00}{1.00}{0.80}{1.00} &2 & 0 & 95.0  & 10.4\% \\          
\quad C-V Contrast& 0.66 & 0.00 & 0.66 & 0.65                                                   & 90\% & 20.00 & \Chart{1.00}{1.00}{1.00}{0.80}{0.70} & 4 & 1 & 92.5 &  16.4\%                             & 52\% & 197.00 & \Chart{1.00}{0.80}{0.70}{0.10}{0.00} &17 & 7 & 91.8  & 13.8\% \\ [0.5em]       

\textbf{Distortion} & & & & & & & & & & & & & & & & & &   \\

\quad D-I Elastic& 0.63 & 0.01 & 0.64 & 0.61                                                 & 98\% & 2.00 & \Chart{1.00}{1.00}{1.00}{1.00}{0.90} & 0 & 1 & 73.3 &  9.6\%                             & 100\% & 0.00 & \Chart{1.00}{1.00}{1.00}{1.00}{1.00} &0 & 0 & 85.1 & 6.6\% \\                 
\quad D-II Pixellate& 0.62 & 0.05 & 0.66 & 0.53                                                & 96\% & 8.00 & \Chart{1.00}{1.00}{0.80}{1.00}{1.00} & 2 & 0 & 84.4 &  14.4\%                               & 98\% & 2.00 & \Chart{1.00}{1.00}{0.90}{1.00}{1.00} &1 & 0 & 96.6  & 8.6\% \\               
\quad D-III Sample & 0.41 & 0.16 & 0.64 & 0.20                                               & 54\% & 213.00 & \Chart{1.00}{0.90}{0.70}{0.10}{0.00} & 21 & 2 & 86.6 &  29.6\%                             & 90\% & 15.00 & \Chart{1.00}{0.90}{0.90}{0.70}{1.00} &5 & 0 & 75.8  & 15.2\% \\             
\quad D-IV Sharpen& 0.66 & 0.01 & 0.67 & 0.64                                                & 60\% & 255.00 & \Chart{1.00}{0.90}{1.00}{0.10}{0.00} & 20 & 0 & 64.0 &  18.6\%                             & 94\% & 8.00 & \Chart{1.00}{0.90}{1.00}{1.00}{0.80} &3 & 0 & 87.1 &  9.4\% \\ [0.5em]                  

\textbf{Affine Transformations} & & & & & & & & & & & & & & & & & &  \\

\quad E-II Scale & - & - & - & - & 70\% & 45.00 & \Chart{0.90}{0.90}{0.60}{0.70}{0.40} & 5 & 10 & 220.7  & 24.6\% & 46\% & 168.00 & \Chart{1.00}{0.20}{0.80}{0.20}{0.10} &15 & 12 & 123.7  & 15.0\% \\              
\quad E-III Translate & - & - & - & - & 20\% & 120.00 & \Chart{0.80}{0.00}{0.20}{0.00}{0.00} & 29 & 11 & 323.8 &  22.2\%   & 64\% & 58.00 & \Chart{0.90}{0.90}{0.50}{0.40}{0.50} &18 & 0 & 74.4 &  11.4\% \\ [0.5em]            
  
\textbf{Graphic Patterns} & & & & & & & & & & & & & & & & & &  \\

\quad F-I Splatter& 0.65 & 0.01 & 0.66 & 0.64                                                 & 58\% & 122.00 & \Chart{0.80}{0.80}{0.80}{0.50}{0.00} & 16 & 5 & 156.8 & 16.0\%                              & 66\% & 108.00 & \Chart{1.00}{1.00}{0.60}{0.40}{0.30} &14 & 3 & 90.3  & 11.8\% \\          
\quad F-II Dotted lines& 0.66 & 0.01 & 0.66 & 0.65                                                & 92\% & 7.00 & \Chart{1.00}{1.00}{0.90}{0.90}{0.80} & 3 & 1 & 92.6 &  18.6\%                                 & 80\% & 65.00 & \Chart{1.00}{1.00}{0.90}{0.70}{0.40} &5 & 5 & 108.2  & 8.2\% \\        
\quad F-III ZigZag& 0.66 & 0.00 & 0.66 & 0.65                                               & 100\% & 0.00 & \Chart{1.00}{1.00}{1.00}{1.00}{1.00} & 0 & 0 & 77.2 &  17.4\%                          & 100\% & 0.00 & \Chart{1.00}{1.00}{1.00}{1.00}{1.00} &0 & 0 & 91.7 &  5.6\% \\                 
\quad F-IV Canny edges& 0.63 & 0.00 & 0.64 & 0.63                                                & 88\% & 2.00 & \Chart{0.80}{0.90}{0.90}{0.90}{0.90} & 6 & 0 & 90.0 &  19.4\%                                & 100\% & 0.00 & \Chart{1.00}{1.00}{1.00}{1.00}{1.00} &0 & 0 & 85.3 & 7.2\% \\            
\quad F-V Cutout& 0.64 & 0.02 & 0.66 & 0.60                                                 & 84\% & 43.00 & \Chart{1.00}{0.90}{1.00}{0.80}{0.50} & 8 & 0 & 123.4 & 12.2\%                         & 66\% & 33.00 & \Chart{0.80}{0.90}{0.60}{0.50}{0.50} &15 & 2 & 99.8 &  14.6\% \\ [0.5em]              

\textbf{Color/Tone Adjustments} & & & & & & & & & & & & & & & & & &  \\

\quad G-I False color& 0.50 & 0.09 & 0.58 & 0.34                                            & 24\% & 153.00 & \Chart{0.00}{0.90}{0.00}{0.00}{0.30} & 38 & 0 & 40.8 &  14.4\%                               & 82\% & 92.00 & \Chart{0.30}{1.00}{1.00}{0.80}{1.00} &5 & 4 & 89.5  & 12.4\% \\            
\quad G-II Phase scrambling& 0.50 & 0.16 & 0.66 & 0.23                                           & 56\% & 113.00 & \Chart{0.90}{0.80}{0.70}{0.20}{0.20} & 19 & 3 & 44.3 &  33.4\%                           & 46\% & 213.00 & \Chart{1.00}{0.90}{0.30}{0.10}{0.00} &21 & 6 & 49.0  & 15.2\%  \\     
\quad G-III Histogram eq.& 0.65 & 0.01 & 0.66 & 0.64                                          & 52\% & 32.00 & \Chart{0.70}{0.70}{0.50}{0.40}{0.30} & 6 & 18 & 197.5  & 49.0\%                              & 74\% & 28.00 & \Chart{1.00}{0.80}{0.70}{0.60}{0.60} &13 & 0 & 75.0 & 9.6\% \\  
\quad G-IV White balance& 0.66 & 0.00 & 0.66 & 0.66                                           & 94\% & 8.00 & \Chart{1.00}{1.00}{1.00}{0.90}{0.80} & 0 & 3 & 91.9 &  29.6\%                                 & 92\% & 2.00 & \Chart{0.90}{0.90}{0.90}{1.00}{0.90} &4 & 0 & 92.2 &  8.0\% \\             
\quad G-V Greyscale& 0.66 & 0.02 & 0.67 & 0.62                                            & 66\% & 173.00 & \Chart{1.00}{0.90}{0.90}{0.50}{0.00} & 13 & 4 & 143.8 &  32.0\%                             & 88\% & 17.00 & \Chart{1.00}{1.00}{0.90}{0.80}{0.70} &6 & 0 & 85.5  & 11.8\% \\          
\quad G-VI Saturation inc.& 0.65 & 0.01 & 0.66 & 0.64                                           & 68\% & 72.00 & \Chart{0.90}{0.90}{0.80}{0.50}{0.30} & 11 & 5 & 118.8 &  22.4\%                                   & 46\% & 258.00 & \Chart{1.00}{1.00}{0.00}{0.30}{0.00} &0 & 27 & 131.5 & 6.2\% \\       
\quad G-VIb Saturation dec.& 0.64 & 0.03 & 0.66 & 0.60                                           & 52\% & 187.00 & \Chart{0.90}{0.80}{0.80}{0.10}{0.00} & 15 & 9 & 103.6 &  17.6\%                                    & 88\% & 7.00 & \Chart{0.90}{1.00}{0.80}{0.90}{0.80} &6 & 0 & 91.7 & 5.8\%\\         
\quad G-VII Posterize& 0.65 & 0.02 & 0.66 & 0.62                                          & 82\% & 32.00 & \Chart{0.90}{0.90}{0.90}{0.90}{0.50} & 4 & 5 & 108.1 &  40.2\%                           & 86\% & 28.00 & \Chart{1.00}{1.00}{0.90}{0.80}{0.60} &7 & 0 & 78.8 & 14.6\% \\              

\bottomrule

\end{tabular}
}
\end{table*}

\begin{table*}[t]
\centering

\caption{RQ\textsubscript{2}: Average metrics for original and extended models on testing datasets.}
\label{tab:rq2}

\resizebox{\textwidth}{!}{%

\begin{tabular}{lcclcccccccccc}

\toprule

\multicolumn{3}{c}{Semantic Segmentation} & \multicolumn{11}{c}{LK/ACC} \\
\cmidrule(r){1-3} \cmidrule(r){4-14}

&&&& \multicolumn{6}{c}{Overall} & \multicolumn{4}{c}{Non-failing} \\
\cmidrule(r){5-10} \cmidrule(r){11-14}

& \multicolumn{2}{c}{Average IoU} 
& 
& \multicolumn{2}{c}{Success rate (\%)} 
& \multicolumn{2}{c}{\# OR} 
& \multicolumn{2}{c}{\# OT} 
& \multicolumn{2}{c}{Time (s)} 
& \multicolumn{2}{c}{Jitter (\%)} \\

\cmidrule(r){2-3} \cmidrule(r){5-6} \cmidrule(r){7-8} \cmidrule(r){9-10} \cmidrule(r){11-12} \cmidrule(r){13-14}

Weather & Original & Extended & Weather & N & FT & N & FT & N & FT & N & FT & N & FT \\

\midrule

nominal & 0.663 & 0.718 & nominal & 64 & 78 & 5 & 3 & 0 & 0 & 28.30 & 34.90 & 2.32 & 5.20 \\ [0.5em]

fog & 0.365 & 0.520 & fog & 14 & 57 & 12 & 6 & 0 & 0 & 24.50 & 26.60 & 4.18 & 3.93 \\
morning & 0.667 & 0.709 & dawn & 57 & 64 & 6 & 5 & 0 & 0 & 73.20 & 24.65 & 2.56 & 4.47 \\
overcast & 0.669 & 0.707 & dark/overcast & 5 & 85 & 7 & 2 & 0 & 0 & 37.25 & 24.40 & 2.66 & 3.33 \\
rain & 0.385 & 0.650 & rain & 42 & 78 & 8 & 3 & 0 & 0 & 57.05 & 26.60 & 3.08 & 3.35 \\
sunset & 0.685 & 0.748 & moonshine & 21 & 78 & 10 & 3 & 1 & 0 & 98.35 & 24.65 & 2.68 & 3.98 \\
-- & -- & -- & snow & 35 & 78 & 9 & 3 & 0 & 0 & 48.85 & 27.10 & 3.08 & 4.39 \\
-- & -- & -- & starry & 21 & 78 & 10 & 3 & 1 & 0 & 98.15 & 26.90 & 2.78 & 3.29 \\

\bottomrule

\end{tabular}
}
\end{table*}

\subsubsection{RQ\textsubscript{2}}

\head{Semantic Segmentation Dataset Augmentation}
We introduce controlled perturbations to the images from the dataset \textit{Augmentation set} (\autoref{sec:dataset}), utilizing the perturbation types identified in \textbf{RQ\textsubscript{1}} at maximum intensity. 

After generating the perturbed images, we use them to perform fine-tuning of the \segformer model, which was trained on the original vKITTI dataset in nominal conditions, by executing one epoch of training.
To evaluate the impact of this augmentation, we test the fine-tuned model on dataset \textit{Testing set} under both nominal conditions (\textit{Testing set (N}) and simulated weather effects (\textit{Testing set (W}). The goal is to measure whether the fine-tuned model exhibits improved generalization and robustness compared to the original, unmodified model. The same evaluation metric of \textbf{RQ\textsubscript{1}} (i.e., IoU) will be used to compare the model's performance on both nominal and perturbed data.

\head{LK/ACC Online Continuous-learning}
We employ a hybrid control system consisting of a pure-pursuit controller for steering and a PID controller for throttle. The pure-pursuit controller calculates the steering angle to keep the vehicle aligned with the road by following predefined waypoints, while the PID controller regulates the throttle based on an expected speed and the car's relative distance to the center of the target lane.
If the car deviates far from the center of the lane, the PID controller reduces the throttle to slow the vehicle down, allowing it to regain control and move back towards the center. This helps to prevent failures in challenging situations. If the car is close to the center of the lane, the throttle is adjusted to match the expected speed parameter, to obtain a vehicle speed consistent with the target speed.
We first deploy the pure-pursuit/PID combo on the \textit{Training roads} set under nominal conditions. This enables us to gather accurate ground truth data that represents good, failure-free driving behaviour. Once the data collection is complete, we train the \davetwo model using this nominal dataset. Hereafter, this model will be referred to as \textit{\davetwo(N)}.

To establish the baseline for how well the model performs without additional training on perturbed scenarios, we evaluate \textit{\davetwo(N)}'s performance on a set of test roads that differ in topology from the training set (\textit{Testing roads\textsubscript{RQ2}}), both in nominal weather conditions and under real weather scenarios in the Udacity simulator. To observe both generalization and robustness improvements, the set of 15 roads (\textit{Testing roads\textsubscript{RQ2}}) has been designed so that \textit{\davetwo(N)} succeeds in 10 out of 15 scenarios in nominal conditions. These tests form the basis for assessing the model's performance without exposure to perturbed environments.
Next, we apply the image perturbations identified in \textbf{RQ\textsubscript{1}} across randomly generated roads, using \tool to create two new random roads (\textit{Fine-tuning roads}) that introduce different driving challenges from those encountered during the nominal evaluation. These random roads are designed to test the model's robustness under varied and unforeseen conditions. During this phase, we apply five intensity levels for each type of image perturbation. While running \textit{\davetwo (N)} to evaluate its robustness under perturbed environments, the pure-pursuit/PID expert driver operates in \textit{shadow mode}, continuously collecting ground truth data.

With the additional perturbed data collected from the randomly generated roads, we conduct one epoch of fine-tuning on the \davetwo model using this new dataset, obtaining \textit{\davetwo(FT)}. 
Finally, we re-evaluate the \textit{\davetwo(FT)} model on both nominal roads and real-world weather conditions in the Udacity simulator. To measure the impact of continuous learning, we compare \textit{\davetwo(N)} and \textit{\davetwo(FT)} models' behaviour using the same metrics used to answer \textbf{RQ\textsubscript{1}}.

\subsection{Results}

\head{RQ\textsubscript{1} (effectiveness)}
\autoref{tab:RQ1} shows our effectiveness results, for both ADAS. 
Concerning semantic segmentation, The left side of \autoref{tab:RQ1} presents the SegFormer model’s performance on the \textit{Testing set (N)} under nominal and perturbed conditions, reporting the average per-class IoU (i.e., the average of the IoU for each class) calculated at each of the five intensity levels. We use the average (Avg.), standard deviation (Std.), maximum (Max.), and minimum (Min.) statistics derived from these values to reflect overall performance and variability. 

The Avg. IoU gives an overall assessment, while Std. shows variability in performance based on intensity. Max. and Min. IoU show the least and the most effective perturbation intensities respectively. Perturbations A-III and D-III were the most disruptive, with Avg. IoUs of $0.41$ (-38\%), closely followed by A-I at $0.43$ (-35\%). In contrast, perturbations like B-V, C-IV, C-V, and the Graphic patterns (F) category had little to no effect, as indicated by consistently low Std. and high Avg. values. 
At the highest intensities, D-III and A-III reduced IoU to $0.20$ (-70\%) and $0.21$ (-68\%), respectively, while A-I, G-II, and others produced minimum IoUs below $0.35$ (-47\%). Perturbations with high Std. values, such as G-II, showed a wide variance in their impact, with disruption ranging from 0\% to 65\%, depending on intensity.

Concerning LK/ACC, the right side of \autoref{tab:RQ1} shows the \davetwo model effectiveness under various perturbations in the Udacity and Donkey Car simulators. We report the average success rate, standard deviation, completion rate trends over the five intensities as a histogram, and failure types---either Out of Road (OR) or Out of Time (OT)---across five intensity levels. For non-failing scenarios, execution time and driving jitter are provided to assess the impact on throttle and steering. 
Eight perturbations had minimal impact, reducing success rates by less than 10\% in both simulators (e.g., A-II, A-III, B-I, B-IV). Five others reduced success rates by less than 20\%. Nine perturbations affected simulators differently, reducing success rates by less than 20\% in Donkey Car but by more than 20\% in Udacity (e.g., A-IV, C-I, D-III). Perturbation E-II, G-II, and G-VI had the largest effect in Donkey Car, lowering success rates to 46\%, while E-III, G-I, and C-II were most disruptive in Udacity, reducing success rates to as low as 20\%. 

Failure types were mostly OR, but some perturbations, like A-I and A-IV, caused more OT failures, especially in Udacity. Perturbations causing OT failures generally increased execution time, while those causing OR failures reduced it, sometimes significantly, as seen with C-I, C-II, and C-III in Udacity. Driving quality also varied, with higher driving jitter in Udacity, indicating a less stable model. For this metric, the most disruptive perturbations were D-III, E-II, and G-VII for D, and A-I, A-IV, and G-VIII for Udacity.

Finally, the driving jitter is significantly higher in Udacity, indicating less stable driving. In Donkey Car, the most disruptive perturbations are D-III, E-II, F-V, G-II, and G-VII, while in Udacity, they are A-I, A-IV, B-II, G-III, and G-VIII, with no clear overlap between domains.

\begin{tcolorbox}
\textbf{RQ\textsubscript{1} (effectiveness):} For both ADAS tasks, most image perturbations impact the robustness, though the effects of the same perturbation type vary across different tasks and ADAS models. For the semantic segmentation task, the most significant impact (-70\%) was observed with the sample (D-III) perturbation, while in the LK/ACC task the more robust model was most affected (-54\%) by phase-scrambling (G-II).
\end{tcolorbox}

\head{RQ\textsubscript{2} (generalization)}
\autoref{tab:rq2} shows the generalization results. 
Concerning semantic segmentation, the leftmost section of \autoref{tab:rq2} details the effectiveness of the SegFormer model, trained on the \textit{Training set}, either in its original form (original) or after fine-tuning with image perturbations from the \textit{Augmentation set} (extended). The model's performance is tested on images from both the \textit{Testing set (N)} (nominal conditions) and the \textit{Testing set (W)} (weather domains). For each scenario, we report the average per-class Intersection over Union (IoU), allowing a direct comparison of the model's performance between the original and extended versions. 

The results show that fine-tuning the SegFormer model with augmented data improves its effectiveness across all weather conditions, particularly in challenging environments like fog and rain, as the IoU increases from $0.36$ to $0.52$ (44\%) and $0.38$ to $0.65$ (71\%), respectively. Nominal conditions also see an improvement from $0.66$ to $0.78$ (18\%), with moderate gains in morning, overcast, and sunset scenarios. 

Concerning LK/ACC, the evaluation results in the right section of \autoref{tab:rq2} show that the effectiveness of \davetwo, initially trained on the \textit{Training roads} (\textit{\davetwo(N)}) and then fine-tuned on the \textit{Fine-tuning roads} (\textit{\davetwo(FT)}). Each row provides the evaluation of the 15 \textit{Testing roads\textsubscript{RQ2}}, both under nominal conditions (row 1) and simulator-based weather conditions.
The table compares the model effectiveness using the success rate (with 100\% representing success on all 15 roads) and reports the number of failures, categorized by type: Out of Road and Out of Time. For non-failing scenarios, we also provide the average execution time and driving jitter.

\davetwo(FT) shows a significant improvement in both success rates and reduction in OR failures compared to \davetwo(N) across all weather conditions. In nominal conditions, the success rate increases to $78\%$ from an initial $64\%$, while the number of failures decreases from $5$ to $3$, indicating a model that generalizes better to new roads. 

In foggy conditions, the success rate increases from $14\%$ to $57\%$, with OR failures decreasing from $12$ to $6$. For dawn, the success rate improves from $57\%$ to $64\%$, and OR failures drop from $6$ to $5$. In dark/overcast conditions, the success rate increases from $5\%$ to $85\%$, and OR failures fall from $7$ to $2$. In rainy conditions, the success rate rises from $42\%$ to $78\%$, with OR failures dropping from $8$ to $3$. Similarly, in moonshine, the success rate increases from $21\%$ to $78\%$, while OR failures reduce from $10$ to $3$. In snow, the success rate improves from $35\%$ to $78\%$, and OR failures drop from $9$ to $3$. Finally, in starry conditions, the success rate climbs from $21\%$ to $78\%$, with OR failures decreasing from $10$ to $3$. 

In terms of OT failures, only moonshine and starry weather caused one OT failure each, which have been both mitigated during fine-tuning.
Driving jitter shows a slight increase in most weather conditions after fine-tuning. In dawn, jitter rises from 2.56\% to 4.47\%; in snow, it increases from 3.08\% to 4.39\%. The increase in jitter is generally minimal, with the exception of fog, where it shows a slight improvement, decreasing from 4.18\% to 3.93\%.

\begin{tcolorbox}
\textbf{RQ\textsubscript{2} (generalization):} 
For both ADAS tasks, fine-tuning the DNN using image perturbations, improves the ADAS effectiveness on unseen, simulated, weather domains, while retaining the original capabilities on nominal scenarios.
\end{tcolorbox}

\section{Discussion}\label{sec:discussion}

\subsection{Effectiveness (RQ\textsubscript{1})}

Our study shows that image perturbations significantly impact the performance of both modular (semantic segmentation) and end-to-end (LK/ACC) ADAS, with varying effectiveness based on perturbation type and intensity. In the offline evaluation of the SegFormer model using the vKITTI dataset, perturbations like D-III Sample and A-III Impulse noise notably degraded performance, reducing IoU from $0.66$ to below $0.21$ at higher intensities. This confirms the vulnerability of vision-based models to visual distortions, even with advanced architectures like transformers. In contrast, perturbations such as Histogram equalization (G-III), White balance (G-IV), and all Graphic patterns (F) had little to no impact, suggesting a greater robustness to global adjustments or artificial patterns.

For the end-to-end LK/ACC system, the perturbations caused more failures in the Udacity simulator compared to the Donkey Car simulator, likely due to the
distinct car dynamics between the simulators as Udacity utilizes wheel friction to move the car, whereas Donkey Car employs a kinematic model that directly translates the car's position based on inputs and current movement.
Perturbation False color (G-I), for example, resulted in an average success rate of only 24\% in the former, while 82\% in the latter. This highlights the importance of employing different testing environments and simulation platforms for cross-validating research results in ADAS testing, as DNN models behave differently between simulators~\cite{biagiola2023better,AminiFlaky2024,borg}. 

An interesting finding of our study is that perturbations impacted the outputs of the ADAS differently. 
Most OR failures were triggered by steering errors of the LK system resulting from visual distortions, while perturbations like Scale (E-II) and Saturation (G-VI) caused OT failures of the ACC system.

\subsection{Generalization (RQ\textsubscript{2})}

Fine-tuning the semantic segmentation model with perturbation-augmented data significantly improved performance across all weather conditions, particularly in fog and rain, where IoU increased from $0.36$ to $0.52$ and from $0.38$ to $0.65$. This shows that even common, arguably non-realistic, perturbations enhance resilience to more naturalistic real-world environmental changes. The robustness in nominal conditions also improved, asserting that our retraining pipeline increased the \textit{overall} robustness of the ADAS rather than overfitting it to the new conditions.

Similarly, fine-tuning LK/ACC through continuous learning with real-time perturbations increased the success rates in all weather scenarios, with an increase of up to $80\%$ in the most challenging condition (i.e., dark/overcast skybox), with fewer OT failures. However, our findings also indicated an increase in driving jitter, resulting in a decrease in control smoothness for the ADAS post-retraining. These results suggest that future work for system-level robustness testing of ADAS should be directed toward balancing both functional and non-functional requirements, as achieving enhanced robustness in novel conditions should not come at the cost of reduced steering precision and a less smooth driving experience. 

\subsection{Threats to Validity}

\subsubsection{Internal validity}

Several factors may affect the internal validity of our study, particularly in the design and execution of the experiments. The use of two simulators with differing car dynamics and distinct LK/ACC models could introduce variations in performance unrelated to image perturbations. To account for this, we report nominal model performance in each evaluation step.

Our computational benchmarks for real-time feasibility were conducted on an Apple M1 processor. While consumer-grade hardware was chosen to reflect practical applications, differences in hardware specifications could affect execution times and perturbation feasibility in other setups.

We used IoU for semantic segmentation and success rates for LK/ACC as primary metrics, aggregated across perturbation intensity levels and scenarios. This aggregation might overlook specific effects at different intensities, which is why full experimental logs are provided in the replication package for more detailed analysis.

\subsubsection{External Validity}

Our system-level experiments were limited to two simulators and focused on Level 2 ADAS, which may limit the applicability of our findings to other simulators and higher levels of autonomy (e.g., Level 3 and Level 4 ADAS). However, studies have shown that there are dozens of simulation platforms available, both commercially and open-source~\cite{Li_2024,10174078}, with no consolidated omni-comprehensive solution. While our study shows that the magnitude and occurrence of failures do change across simulators, this choice does not undermine the core insights of our study. 

Our evaluations were conducted in simulated environments, which may not fully replicate the complexities of real-world driving conditions. While our perturbations mimic plausible visual distortions (e.g., weather conditions, noise, and lighting changes) and the simulated weather domains do represent realistic phenomena, real-world driving environments involve more complex scenarios and real-world sensors are subject to hardware-specific noise and physical degradation that cannot be fully simulated. As a result, the robustness gains observed  using simulation may not directly translate to improvements in real-world environments.

\subsubsection{Reproducibility} 

The entire pipeline used to obtain the results discussed in this work, including model training, our library \tool, metric calculations, and results, is available and can be reproduced~\cite{tool}.
\section{Related work}\label{sec:related-work}

\subsection{Model-level Studies}

Most research effort has been directed towards robustness testing of image classifiers~\cite{2025-Maryam-ICST}, proposing benchmark datasets of corrupted images, such as ImageNet-C~\cite{hendrycks2019benchmarking}, and \mbox{MNIST-C}~\cite{mu2019mnistc} or augmentation techniques to enhance the diversity of training data, such as RandAugment~\cite{cubuk2019randaugment} and AugMix~\cite{hendrycks2020augmix}. 
In contrast, we focus on ADAS, where a lack of DNN robustness can lead to safety-critical failures.

In the ADAS domain, model-level testing efforts predominantly target a restricted range of perturbations. Tools such as DeepXplore~\cite{deepxplore} use neuron coverage metrics to uncover misbehaviours in DNNs by applying only a few image perturbations types, such as lighting effects and occlusion by single or multiple small rectangles.
DeepTest~\cite{deeptest} employs perturbation types like rotation, translation, and shear, which are unlikely to induce realistic ADAS misbehaviours since they do not correspond to realistic driving scenarios. 

Differently, in our work, we retrieved the most complete list of image perturbations from existing literature. We filtered out those that do not produce valid driving images, as well as those that are too computationally expensive for system-level testing. Furthermore, our research extends beyond robustness evaluation by investigating the potential of common image perturbations to enhance the generalizability of ADAS during domain adaptation and retraining campaigns.

\subsection{System-level Studies}

Among the adversarial attack techniques, Wu et al.~\cite{dataAugment2020Liu} developed a real-time adversarial attack on an end-to-end driving model, which can force the vehicle to deviate from its designated lane. Similarly, Yoon et al.~\cite{yoon2023learning} introduced an online image attack framework, which utilizes a binary decision boundary to decide when to launch attacks.

Other research leveraged image perturbations to create efficient runtime performance prediction modules for ADAS, applying these perturbations at a system level to evaluate potential errors rather than to execute driving maneuvers. 
For example, Luan et al.~\cite{Luan2023Efficient} introduce nine specific perturbations to each input frame, and calculates an anomaly score by comparing the driving commands from perturbed and unperturbed images. MarMot~\cite{ayerdi2023metamorphic} implemented a runtime monitoring framework using five domain-specific metamorphic relations to influence the ADAS output, allowing for the generation of confidence scores by comparing the predictions from original and altered images at each frame. DeepManeuver~\cite{DeepManeuver} proposes a state-aware robustness testing framework, using road perturbations to expose failures in end-to-end driving models.

Our study complements these efforts by focusing on both modular and end-to-end ADAS, applying common perturbations during system-level evaluation using two simulators. While existing works only target adversarial attacks, specific perturbation types, and prediction reliability, our research systematically applies a broad range of image perturbations, road tracks, and multiple  metrics, including success rates, failure types, and driving jitter, and domain generalization.

\section{Conclusions and Future Work}\label{sec:conclusions}

Our study systematically evaluates the robustness of vision-based ADAS for perception tasks using 32 image perturbations from existing literature and assessing their impact at both the model level (semantic segmentation) and system level (LK/ACC). 
We found that perturbations such as Phase scrambling were particularly disruptive, significantly reducing performance across domains, while others such as ZigZag patterns had minimal effects. This evaluation highlights that perturbations can affect ADAS differently depending on whether they target offline perception tasks or real-time driving scenarios. Our findings further demonstrate that applying perturbation-based data augmentation and continuous learning improves ADAS robustness, particularly in adverse weather conditions, such as fog and rain. This approach not only increased robustness to real-world environmental changes but also improved generalization on new scenarios under nominal conditions, emphasizing the value of perturbations in enhancing perception models' robustness.

In our future work we will investigate white-box adaptive perturbations and also consider extending the study to additional ADAS and autonomous driving stacks. 

\section*{Acknowledgements}
\addcontentsline{toc}{section}{Acknowledgements}
This research was funded by the Bavarian Ministry of Economic Affairs, Regional Development and Energy.

\balance
\bibliographystyle{IEEEtran}
\bibliography{paper}

\end{document}